\documentclass{article}
\usepackage{amssymb}
\usepackage{graphics}

\newtheorem{theorem}{Theorem}
\newtheorem{acknowledgement}[theorem]{Acknowledgement}

\begin{document}

\title{Effective resonant transitions in quantum optical systems: kinematic
and dynamic resonances}
\author{A.B. Klimov$^{1}$, I. Sainz$^{2}$ \\
$^{1}$Departamento de F\'{\i}sica, Universidad de Guadalajara, \\
Revoluci\'{o}n 1500, Guadalajara 44420, M\'{e}xico.\\
$^{2}$Centro Universitario de los Lagos, Universidad de Guadalajara,\\
Enrique Diaz de Le\'{o}n s/n, Lagos de Moreno 47460, M\'{e}xico.}
\maketitle

\begin{abstract}
We show that quantum optical systems preserving the total number of
excitations admit a simple classification of possible resonant transitions
(including effective), which can be classified by analizying the free
Hamiltonian and the corresponding integrals of motion. Quantum systems not
preserving the total number of excitations do not admit such a simple
classification, so that an explicit form of the effective Hamiltonian is
needed to specify the allowed resonances. The structure of the resonant
transitions essentially depends on the algebraic propereties of interacting
subsystems.
\end{abstract}

\section{Introduction}

In a direct analogy with classical mechanics, composed systems in quantum
optics\ (which describe interaction between several subsystems) can be
divided into two classes: 1) Systems which possess a necessary number of
integrals of motion, so that, the classical counterpart is an integrable
system; 2) Systems that do not admit a sufficient number of integrals of
motions, so that, the classical counterpart is a non-integrable system \cite%
{caos}. A quantum system can have basically two types of integrals of
motion: a) \textit{Kinematic }integrals, which do not depend on the kind of
interaction between subsystems, as for instance, the total number of atoms;
b) \textit{Dynamic} integrals which are related to the particular form of
interaction between subsystems, as, for instance, a number of excitations
preserved in some transitions between the energy levels of the subsystems.

Typical for quantum optical systems dipole-like interactions between two
subsystems ($X$ and $Y$) can be described with a generic multichannel
Hamiltonian of the following form:%
\begin{equation}
H=\sum\limits_{j}\omega _{j}X_{0}^{j}+\sum\limits_{k}\Omega
_{k}Y_{0}^{k}+\sum\limits_{j,k}g_{jk}\left( X_{-}^{j}+X_{+}^{j}\right)
\left( Y_{-}^{k}+Y_{+}^{k}\right) ,  \label{H1}
\end{equation}%
where the two first terms represent the free Hamiltonians of the subsystems,
so that the frequencies $\omega _{j},\Omega _{k}\geq 0$, and the last term
describes the interaction between them. The operators $X_{0,\pm }^{j}$, $%
Y_{0,\pm }^{j}$, $[X_{0,\pm }^{j},Y_{0,\pm }^{j}]=0$, are usually elements
of some deformed algebra \cite{Karassiov}, and in particular, satisfy the
ladder commutation relations%
\begin{equation}
\left[ X_{0}^{j},X_{\pm }^{j}\right] =\pm X_{\pm }^{j},\hspace{0.1in}\left[
Y_{0}^{j},Y_{\pm }^{j}\right] =\pm Y_{\pm }^{j}.  \label{escal}
\end{equation}

In the interaction Hamiltonian there are two kinds of terms: of the form $%
X_{-}^{j}Y_{+}^{k}$ and $X_{-}^{j}Y_{-}^{k}.$ It is easy to observe that in
the rotating frame, that is, applying the following unitary transformation%
\[
U=\exp \left( it\left[ \sum\limits_{j}\omega
_{j}X_{0}^{j}+\sum\limits_{k}\Omega _{k}Y_{0}^{k}\right] \right)
\]%
to the Hamiltonian (\ref{H1}), the \textit{counterrotating }terms $\sim
X_{-}^{j}Y_{-}^{k}$ oscillate in time with a frequency $\omega _{j}+\Omega
_{k}$ and the \textit{rotating }terms $\sim X_{-}^{j}Y_{+}^{k}$ oscillate
with a frequency $\omega _{j}-\Omega _{k}$. It is clear that under the
condition $\omega _{j}\approx \Omega _{k}$ the rotating term $%
X_{-}^{j}Y_{+}^{k}$ in (\ref{H1}) is approximately time independent (and
thus, can generate transitions with a probability of one between the energy
levels of the system), meanwhile the counterrotating\textit{\ }term, $%
X_{-}^{j}Y_{-}^{k}$ always oscillates rapidly in the rotating frame, and its
temporal average is zero.

By neglecting the counterrotating terms in the Hamiltonian (\ref{H1}), which
is commonly called the Rotating Wave Approximation (RWA), we arrive at the
Hamiltonian%
\begin{equation}
H=\sum\limits_{j}\omega _{j}X_{0}^{j}+\sum\limits_{k}\Omega
_{k}Y_{0}^{k}+\sum\limits_{j,k}g_{jk}\left(
X_{-}^{j}Y_{+}^{k}+X_{+}^{j}Y_{-}^{k}\right) ,  \label{H3}
\end{equation}%
which admits several dynamic integrals of motion $N_{p}$, generally not
allowed in (\ref{H1}). This implies that the whole representation space of
the system is divided into finite dimensional invariant subspaces, and the
mathematical treatment is essentially simplified.

It is worth noting, that the semiclassical models, when some of the
subsystems are described by $c$-numbers instead of operators, are treated
using essentially the same type of Hamiltonians as (\ref{H1}). The
semiclassical transition in (\ref{H1}) can be done by going to the rotating
frame of the semiclassical system and then, just substituting the transition
operators by some complex numbers. For instance, in the case of single
channel interaction, $j=1$, $k=1$, when the system $Y$ acquires classical
features, the Hamiltonian (\ref{H1}) in the rotating frame corresponding to
the system $Y$ takes the form
\[
H=\omega X_{0}+g\left( X_{-}+X_{+}\right) \left( Y_{-}e^{i\Omega
t}+Y_{+}e^{-i\Omega t}\right) ,
\]%
so that the corresponding semiclassical Hamiltonian is obtained by
substituting $Y_{+}\rightarrow g,Y_{-}\rightarrow g^{\ast },$ giving%
\begin{equation}
H_{sc}=\omega X_{0}+g\left( X_{-}+X_{+}\right) \left( g^{\ast }e^{i\Omega
t}+ge^{-i\Omega t}\right) .  \label{Hcf1}
\end{equation}%
Such Hamiltonians usually appear when a quantum oscillator and/or a
collection of atoms is pumped by an external force \cite{shirley}, \cite%
{Cohen}, \cite{yabuzaki}.

In the RWA-like systems, described by Hamiltonians of the form (\ref{H3}),
the resonance conditions%
\begin{equation}
\omega _{j}\approx \Omega _{k},  \label{res1}
\end{equation}%
means that the term $X_{-}^{j}Y_{+}^{k}$, which explicitly appears in the
Hamiltonian, does not depend on time in an appropriate rotating frame.
Nevertheless, such \textit{explicit resonances} are not the only kind of
resonances, which can be found in the Hamiltonian (\ref{H3}). Usually, the
composed systems admit several types of \textit{implicit resonances} related
to effective transitions, which do not appear in the original Hamiltonian,
between their energy levels. Such effective interactions play important
roles in many physical applications and can be revealed by adiabatic
elimination of slow transitions \cite{adiabatic}. The implicit (effective)
resonances are characterized by their position, $\sum_{j}m_{j}\omega
_{j}\approx \sum_{k}n_{k}\Omega _{k}$, and strength, i.e. in what order of
perturbation expansion they appear for the first time. Although, a generic
system can possess a large number of different types of effective
transitions, all the possible resonance conditions can be classified only by
analyzing the free Hamiltonian and the integrals of motion. Because all the
invariant subspaces are finite-dimensional, there are always a finite number
of different resonances. We will refer to these kinds of resonances as
\textit{kinematic resonances}, which include both explicit and implicit
resonances. In Sec. II we show with the example of atom-field interactions,
that it is possible to classify all the kinematic resonances in a
straightforward way.

The situation is quite different in quantum systems with a lack of integrals
of motion, corresponding to classically non-integrable dynamic systems. In
such systems an infinite number of different resonances arise and \textit{a
priori} it is impossible to determine their position and strength, which
essentially depend not only on the type of interaction but also on the
algebraic properties of each interacting subsystem \cite{sinrwa}, \cite%
{bombeo}, \cite{pla}. In Sec.III we will discuss such \textit{dynamic
resonances} analyzing different models of interaction of quantum and
classical fields with atomic systems.

\section{Kinematic resonances}

\subsection{A simple model}

An important example of kinematic resonances is the Dicke Model \cite{Dicke}%
, this model describes the interaction of a collection of $A$ identical
two-level atoms with a single mode of a quantized field under the Rotating
Wave Approximation. The Hamiltonian that governs this system is given by
\begin{equation}
H=\omega a^{\dagger }a+\omega _{0}S_{z}+g(aS_{+}+a^{\dagger }S_{-}),
\label{HD}
\end{equation}%
where $\omega $ is the field frequency, $\omega _{0}$ is the atomic
frequency, and $S_{z}=\sum\limits_{i=1}^{A}s_{zi},S_{\pm
}=\sum\limits_{i=1}^{A}s_{\pm i}$ are the collective atomic operators, they
represent the atomic inversion, and the transition between the atomic energy
levels respectively, their commutation relations are given by the $su(2)$
algebra,
\begin{equation}
\left[ S_{z},S_{\pm }\right] =S_{\pm },\quad \left[ S_{+},S_{-}\right]
=2S_{z},\quad   \label{su2}
\end{equation}%
and $a^{\dagger },a$ are the creation-annihilation field operators, obeying
the bosonic commutation relations, $[a,a^{\dagger }]=1$.

The Hamiltonian (\ref{HD}) admits two integrals of motion, a kinematic
integral of \ motion, given by the total number of atoms (that is constant
for a closed system), $A=S^{11}+S^{22}$, where $S^{ii}$ is the atomic
population operator for the $i$-th\ atomic energy level, and the dynamic
integral of motion, corresponding to the total number of excitations, has
the form
\begin{equation}
N=S_{z}+a^{\dagger }a,\qquad \lbrack N,H]=0.  \label{N}
\end{equation}

This system is the simplest non trivial example of one-channel quantum
transitions: an absorption of one photon is accompanied by an excitation of
one atomic transition. Using the integral of motion (\ref{N}), the
Hamiltonian (\ref{HD}) can be rewritten as follows,%
\[
H=\omega N+H_{int},
\]%
where the interaction Hamiltonian is%
\begin{equation}
H_{int}=\Delta S_{z}+g(aS_{+}+a^{\dagger }S_{-}),\quad  \label{HD2}
\end{equation}%
and $\Delta =\omega _{0}-\omega $, is the detuning between the field and the
atomic transition frequencies. If the only possible resonant condition, $%
\omega =\omega _{0},$ is held, the interaction Hamiltonian (\ref{HD2}) is
reduced to,
\begin{equation}
H_{int}=g(aS_{+}+a^{\dagger }S_{-}),  \label{HD1}
\end{equation}%
which implies that the atomic transition probability (as a function of time)
oscillates between zero and one.

On \ the other hand, in the far-off resonant (dispersive) limit, $|\Delta
|\gg g\sqrt{\bar{n}}$, the interaction Hamiltonian (\ref{HD2}) is diagonal,

\begin{equation}
H_{int}\simeq \Delta S_{z}+\frac{g^{2}}{\Delta }\left[ (2a^{\dagger
}a+1)S_{z}-S_{z}^{2}\right] ,  \label{Hdeff}
\end{equation}%
so that only the phase of the system evolves due to the appearance of the
so-called dynamic Stark shift terms \cite{stark}, \cite{DD} in the \textit{%
effective} Hamiltonian (\ref{Hdeff}).

\subsection{Generic atom-field interactions}

Let us consider the interaction of a system of $A$ identical $N$-level atoms
of an arbitrary configuration with a single mode of a quantized field of
frequency $\omega $. The Hamiltonian describing this system has the form

\begin{equation}
H=H_{d}+H_{int},  \label{HC1}
\end{equation}%
where%
\begin{equation}
H_{d}=\omega a^{\dagger }a+\sum\limits_{j=1}^{N}E_{j}S^{jj},\hspace{0.1in}%
E_{j}<E_{j+1},  \label{Hdiag}
\end{equation}%
is the free Hamiltonian, and\ $S^{jj}$ are the collective atomic population
operators corresponding to the $j$-th atomic level of energy $E_{j}$, and $%
H_{int}$ is the interaction atom-field Hamiltonian, whose explicit form
depends on the atomic configuration, in the dipole approximation, i.e. only
one-photon transitions are allowed. In this Section we suppose that the
Rotating Wave Approximation is imposed, so that the total number of
excitations in the atom-field system is preserved, and thus, the whole
representation space of this quantum system is divided into
finite-dimensional invariant subspaces.

The Hamiltonian (\ref{HC1}) admits two integrals of motion: a kinematic
integral, given by the total number of atoms:%
\begin{equation}
A=\sum\limits_{i=1}^{N}S^{ii},  \label{IC}
\end{equation}%
and a dynamic integral, corresponding to the total number of excitations in
the system:%
\begin{equation}
N=a^{\dagger }a+\sum\limits_{i=1}^{N}\mu _{j}S^{ii},  \label{ID}
\end{equation}%
where the parameters $\mu _{j}$ depend on the atomic configuration.

Let us note that a \textit{generic} interaction term can be written as
follows
\begin{equation}
f(S^{ii},a^{\dagger }a)\prod\limits_{j=1}^{N-1}\left( S_{+}^{jN}\right)
^{k_{j}}a^{k_{N}},\hspace{0.1in}k_{j}\in Z\ \forall \ j=1,...,N,  \label{TG}
\end{equation}%
where $S_{\pm }^{ij}$ ($S_{+}^{ij}=S^{ij},$ $S_{-}^{ij}=S^{ji}$, $j>i$) are
the atomic transition operators satisfying the $u(N)$ commutation relations,
$\left[ S^{ij},S^{km}\right] =\delta _{im}S^{kj}-\delta _{kj}S^{im}$, and
negative exponents correspond to the Hermitian conjugated operators. The
operational coefficients $f(S^{ii},a^{\dagger }a)$ depend on diagonal atomic
operators and the photon number operator. In what follows we will omit the
coefficient $f(S^{ii},a^{\dagger }a)$, since it leads only to some phase
shifts and does not change the distribution of excitations in the system%
\'{}%
s energy levels. It is worth noting that (\ref{TG}) is not a unique way to
represent a generic interaction term.

Since the Rotating Wave Approximation is imposed, every interaction term
should preserve the total number of excitations, so the condition%
\[
\left[ N,\prod\limits_{j=1}^{N-1}\left( S_{+}^{jN}\right) ^{k_{j}}a^{k_{N}}%
\right] =0,
\]%
is held, and thus, the numbers $k_{j}$ satisfy the following restriction
\begin{equation}
k_{N}=\sum_{j=1}^{N-1}k_{j}\left( \mu _{N}-\mu _{j}\right) .  \label{k}
\end{equation}%
Thus, any admissible interaction can be described as%
\begin{equation}
\prod\limits_{j=1}^{N-1}\left( S_{+}^{jN}\right)
^{k_{j}}a^{\sum\limits_{j=1}^{N-1}k_{j}\left( \mu _{N}-\mu _{j}\right) }.
\label{git}
\end{equation}%
The interaction (\ref{git}) becomes resonant when the atomic transition
energies and the field frequency satisfy the following condition%
\begin{equation}
\sum\limits_{j=1}^{N-1}k_{j}\left( E_{N}-E_{j}-\omega \left( \mu _{N}-\mu
_{j}\right) \right) =0.  \label{resC}
\end{equation}%
The important point here is that interaction terms (\ref{git}) can be
explicitly presented in (\ref{HC1}), or can describe effective interactions,
and thus should be obtained from the original Hamiltonian by adiabatic
elimination of some far-off resonant transitions. So that, the condition (%
\ref{resC}) describes \textit{explicit} resonances if the corresponding
interaction term is present in the original Hamiltonian or \textit{implicit}
resonances, if such interaction is effective. It is clear that the total
number of both explicit and implicit resonances is finite, which is a
consequence of the restriction (\ref{k}) imposed by the Rotating Wave
Approximation.

The number of possible resonances depends on the number of atomic levels, $N$%
, and the total number of atoms, $A$. Note that the resonance condition (\ref%
{resC}) is associated with a vector $\vec{k}=\left( k_{1},...k_{N-1}\right) $%
, in order to not repeat the resonances, we consider only vectors $\vec{k}$\
with coprime components (that is $k_{1},...k_{N-1}$ do not have a common
factor). Then, different vectors $\vec{k}$ \ satisfying the following
condition\textbf{\ }%
\begin{equation}
\max \left\{ \left\vert \sum\limits_{i=1}^{N-1}k_{i}\right\vert
,\left\vert \sum\limits_{i=1,j\neq i}^{N-1}k_{i}\right\vert
_{j=1,...,N-1},....,\left\vert k_{i}\right\vert
_{i=1,...,N-1}\right\} \leq A,  \label{pr}
\end{equation}%
define different resonances.

On the other hand, taking into account the dynamic integral of motion (\ref%
{ID}), one can rewrite the free Hamiltonian (\ref{Hdiag}) as follows,%
\begin{equation}
H_{d}=\omega N+\sum\limits_{j=1}^{N}\left( E_{j}-\mu _{j}\omega \right)
S^{jj}.  \label{Hdiag1}
\end{equation}%
Summing the kinematic integral $-\beta \sum_{i=1}^{N}S^{ii}$ (\ref{IC}) to
the above Hamiltonian, where the constants $\beta =\beta (\vec{n}),\vec{n}%
=(n_{1},..,n_{N})$, are chosen such that,%
\[
\beta (\vec{n})=\sum\limits_{j=1}^{N}n_{j}\left( E_{j}-\mu _{j}\omega
\right) ,\hspace{0.1in}\sum\limits_{j=1}^{N}n_{j}=1,\quad n_{j}\in \mathcal{Z%
},
\]%
we obtain
\begin{equation}
H_{d}=\omega N+\sum\limits_{j=1}^{N}\left( E_{j}-\mu _{j}\omega -\beta (\vec{%
n})\right) S^{jj}.  \label{Hdiag2}
\end{equation}%
It is easy to see that the condition $E_{j}-\mu _{j}\omega -\beta (\vec{n}%
)=0,$ for any fixed $j$ and some values of $\beta (\vec{n})$, enumerates all
the possible resonances (\ref{resC}). Let us take for instance $j=1,$ if $%
E_{1}-\mu _{1}\omega -\beta (\vec{n})=0,$ then%
\begin{equation}
\sum\limits_{j=2}^{N-1}n_{j}\left( E_{N}-E_{j}-\omega \left( \mu _{N}-\mu
_{j}\right) \right) +\left( n_{1}-1\right) \left( E_{N}-E_{1}-\omega \left(
\mu _{N}-\mu _{1}\right) \right) =0,  \nonumber
\end{equation}%
which coincides with (\ref{resC}), when $k_{j}=n_{j}$ for $j=2,...N-1$ and $%
k_{1}=\left( n_{1}-1\right) $.

This means that we can always represent the free Hamiltonian in a way that
all the possible resonance conditions, corresponding to both explicit and
implicit resonances, appear as zeros of coefficients of the atomic
population operators $S^{jj}$ in the free Hamiltonian, after taking out the
integral of motion (\ref{ID}) corresponding to the total number of
excitations.

The above allows us to classify all the possible kinematic resonances:

\begin{enumerate}
\item \textit{Multiphoton resonances: }transitions which involve absorption
and emission of photons: a) simple $n$-photon transitions, described by the
terms $\sim a^{n}S_{+}^{kj}$, with the resonance condition $%
E_{j}-E_{k}\approx n\omega $. The terms with $n=1$ can be present in the
original Hamiltonian and in such case represent explicit resonances.
Multiphoton transitions appear even in a single atom case; b) Collective
atomic transitions, described by the terms $\sim a^{n}S_{+}^{kl}S_{-}^{ji},$
(in general a product of several, up to $A$ atomic transition operators can
appear), with the corresponding resonance condition $\left(
E_{l}-E_{k}\right) -\left( E_{i}-E_{j}\right) \approx n\omega $. It is clear
that such resonances can appear only in multi-atom systems. Note, that if $%
l=i$ (or $k=j$) then we obtain the same resonance condition as in a),
describing an effective process of absorption of $n$ photons with atomic
transition from $k$-th to $j$-th energy levels. Nevertheless, the
corresponding term in the effective Hamiltonian would be multiplied by the
atomic population operator $S^{ll}$, which means that more than one atom is
need to realize such a process.

\item \textit{Virtual} \textit{photon resonances}: atomic transitions
between independent channels caused by the quantum field fluctuations, and
thus existing even when the field is in the vacuum state. Such resonances
appear only when the system has more than one atom and are described by
terms in the form of products of atomic transitions operators. For instance,
the simplest term of this kind (typically appearing in the first order
perturbation expansion) is $S_{+}^{kj}S_{-}^{il}$, $k\neq i$, $j\neq l$,
describes atomic transitions $k\rightarrow j$, $l\rightarrow i$, and the
corresponding resonance condition is
\begin{equation}
\left( E_{j}-E_{k}\right) -\left( E_{l}-E_{i}\right) \approx 0.  \label{2arc}
\end{equation}%
Obviously, such transitions can be realized in a system which consists of at
least two atoms. Note that the term $\sim S_{+}^{lj}S_{-}^{ik},$ which
represents the atomic transition $k\rightarrow i$, $l\rightarrow j$,
satisfies the same resonant condition. More involved interactions, like$\sim
S_{+}^{ij}S_{-}^{kl}S_{-}^{mn}$, can appear in the highest orders of the
perturbation theory. The strength of virtual photon transitions does not
depend on the field intensity in the leading order of the perturbation
expansion.\ \ \ \ \ \ \ \ \ \ \ \ \ \ \ \ \ \ \ \ \ \ \ \ \ \ \ \ \ \ \ \ \
\ \ \ \ \

\item \textit{Photon assisted transitions}: atomic transitions when every
photon emission is accompanied by a simultaneous photon absorption. These
transitions appear only when the atomic configuration contains coherent
channels, similar to lambda-like configurations. The strength of such
interactions depends on the number of photons in the field and the
populations of some atomic energy levels. The simplest terms describing the
photon assisted transition (which appear in the lowest order of the
perturbation theory) have the form: $f(a^{\dagger }a,S^{jj})S_{+}^{kl}$,
corresponding to the resonance condition $E_{l}\approx E_{k}$, where the
transition $k\leftrightarrow l$ is not present in the original Hamiltonian
and $f(a^{\dagger }a,S^{jj})$ is a linear polynomial of the photon number
operator and the atomic population operators.
\end{enumerate}

In the above classification we do not consider interactions corresponding to
powers of terms describing some interactions. For instance, the $n$-photon
transition corresponding to the term $(aS_{+}^{jk})^{n},$ that describes an
absorption of $n$ photons by at least $n$ atoms in the transition $%
j\rightarrow k$, we include in "one photon transitions".

In the case of interaction of an atomic system with classical field the
Rotating Wave Approximation implies that the interaction Hamiltonian can
always be reduced to a time-independent form. In this case, apart from
explicit resonances, several types of implicit (effective) resonant
transitions can take place. It is clear that no interactions similar to
virtual photon resonances can arise. Nevertheless, transitions similar to
multiphoton resonances $\sim S_{+}^{jk},k-j\geq 2$ and photon assisted
resonances $\sim S_{+}^{jk}$, where the transition $k\leftrightarrow j$ does
not exist in the initial Hamiltonian, actually appear.

\subsection{Four level diamond configuration atoms}

As an example of the kinematic resonances we will study the interaction of $A
$ four level diamond configuration atoms with a single mode of a quantized
field under RWA. The Hamiltonian governing the evolution of this system has
the form%
\begin{eqnarray}
H &=&\omega a^{\dagger }a+\sum\limits_{j=1}^{4}E_{j}S^{jj}+g_{1}\left(
aS_{+}^{12}+a^{\dagger }S_{-}^{12}\right) +g_{2}\left(
aS_{+}^{13}+a^{\dagger }S_{-}^{13}\right)   \label{H4lev} \\
&&+g_{3}\left( aS_{+}^{24}+a^{\dagger }S_{-}^{24}\right) +g_{4}\left(
aS_{+}^{34}+a^{\dagger }S_{-}^{34}\right) .  \nonumber
\end{eqnarray}%
This Hamiltonian describes four one-photon atomic transitions, ($%
1\leftrightarrow 2$, $1\leftrightarrow 3)$, ($2\leftrightarrow 4$, $%
3\leftrightarrow 4)$, gathered in two pairs of coherent quantum channels.
The corresponding resonance conditions (explicit resonances) are $%
E_{2}-E_{1}\approx \omega $, $E_{3}-E_{1}\approx \omega $, $%
E_{4}-E_{2}\approx \omega $, and $E_{4}-E_{3}\approx \omega $. To find all
the other (implicit) resonances we will follow the method outlined in
previous Subection. The dynamic integral of motion for this system is
\[
N=a^{\dagger }a+S^{44}-S^{11},
\]%
so that the coefficients $\mu _{j}$ in (\ref{ID}), are $\mu _{1}=-1$, $\mu
_{2}=\mu _{3}=0$, $\mu _{4}=1$. Substituting these $\mu _{j}$ into (\ref%
{resC}) we obtain all the possible resonance conditions:
\begin{equation}
k_{1}\left( E_{4}-E_{1}-2\omega \right) +k_{2}\left( E_{4}-E_{2}-\omega
\right) +k_{3}\left( E_{4}-E_{3}-\omega \right) \approx 0,  \label{resI}
\end{equation}%
$k_{1},k_{2},k_{3}\in \mathcal{Z},$ and the corresponding (effective)
interactions are%
\[
\left( S_{+}^{14}\right) ^{k_{1}}\left( S_{+}^{24}\right) ^{k_{2}}\left(
S_{+}^{34}\right) ^{k_{3}}a^{k_{3}+k_{2}+2k_{1}}.
\]%
The vector $\vec{k}=\left( k_{1},k_{2},k_{3}\right) $ associated with
possible resonances should satisfy the condition (\ref{pr})%
\[
\max \left\{ \left\vert \sum\limits_{i=1}^{3}k_{i}\right\vert
,\left\vert k_{i}+k_{j}\right\vert _{i\neq j},\left\vert
k_{i}\right\vert \right\} \leq A,\ \ \ \ i,j=1,2,3.
\]

For a single atom we have the vectors $\left( 1,0,0\right) $, $\left(
0,1,0\right) $, $\left( 0,0,1\right) $, $\left( 1,-1,0\right) $, $\left(
1,0,-1\right) $, $\left( 0,1,-1\right) $, which correspond to the following
resonance conditions (interactions): $E_{4}-E_{1}\approx 2\omega $ $\left(
a^{2}S_{+}^{14}\right) $, $E_{4}-E_{2}\approx \omega $ $\left(
aS_{+}^{24}\right) $, $E_{4}-E_{3}\approx \omega $ $\left(
aS_{+}^{34}\right) $, $E_{2}-E_{1}\approx \omega $ $\left(
aS_{+}^{12}\right) $, $E_{3}-E_{1}\approx \omega $ $\left(
aS_{+}^{13}\right) $, $E_{3}\approx E_{2}$ $\left( S_{+}^{23}\right) $,
respectively. For two atoms there there are 15 other vector apart from 6
shown above.

\begin{figure}[h]
\begin{center}
\scalebox{0.2}{\includegraphics{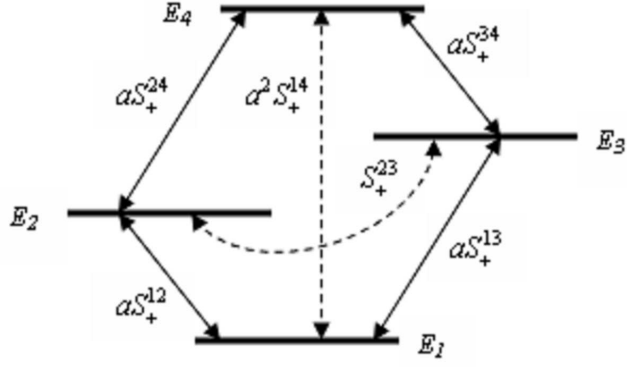}}
\caption{Interactions diagram for a single four level atom of diamond
configuration. Solid line: explicit resonances. Dashed line: Implicit
resonances: two photon transition, photon assisted transition. \label{figura1}}
\end{center}
\end{figure}

For a better understanding of the nature of the implicit resonances, we find
the first order effective Hamiltonian considering that all the interactions
appearing in the initial Hamiltonian (\ref{H4lev}) are far from resonance.
Following the method outlined in the Appendix, we apply to the Hamiltonian (%
\ref{H4lev}) the following sequence of unitary transformations
\[
H_{eff}=U_{4}U_{3}U_{2}U_{1}HU_{1}^{\dagger }U_{2}^{\dagger }U_{3}^{\dagger
}U_{4}^{\dagger },
\]%
where $U_{i}=\exp \left( \varepsilon _{i}A_{i}\right) ,$
\begin{eqnarray*}
A_{1} &=&aS_{+}^{12}-a^{\dagger }S_{-}^{12},\ \ \ \
A_{2}=aS_{+}^{13}-a^{\dagger }S_{-}^{13}, \\
A_{3} &=&aS_{+}^{24}-a^{\dagger }S_{-}^{24},\ \ \ \
A_{4}=aS_{+}^{34}-a^{\dagger }S_{-}^{34},
\end{eqnarray*}%
and the small parameters $\varepsilon _{i}\ll 1$\ are given by%
\begin{eqnarray*}
\varepsilon _{1} &=&\frac{g_{1}}{E_{2}-E_{1}-\omega },\ \ \ \ \varepsilon
_{2}=\frac{g_{2}}{E_{3}-E_{1}-\omega }, \\
\varepsilon _{3} &=&\frac{g_{3}}{E_{4}-E_{2}-\omega },\ \ \ \ \varepsilon
_{4}=\frac{g_{4}}{E_{4}-E_{3}-\omega },
\end{eqnarray*}%
we obtain, in the first order on small parameters, the following effective
Hamiltonian%
\begin{eqnarray}
H &=&\omega a^{\dagger }a+\sum\limits_{j=1}^{4}E_{j}S^{jj}+\Phi (a^{\dagger
}a,S^{jj})+g_{2}\varepsilon _{1}\left[ S_{+}^{23}\left( S^{11}+a^{\dagger
}a+1\right) +h.c.\right]   \nonumber \\
&&+g_{4}\varepsilon _{3}\left[ S_{+}^{23}\left( S^{44}-a^{\dagger }a\right)
+h.c.\right] -\left( g_{3}\varepsilon _{1}+g_{4}\varepsilon _{2}\right)
\left( a^{2}S_{+}^{14}+h.c.\right)   \nonumber \\
&&+g_{3}\varepsilon _{1}\left( S_{+}^{24}S_{-}^{12}+h.c.\right)
+g_{3}\varepsilon _{2}\left( S_{+}^{24}S_{-}^{13}+h.c.\right)   \nonumber \\
&&+g_{4}\varepsilon _{1}\left( S_{+}^{12}S_{-}^{34}+h.c.\right)
+g_{4}\varepsilon _{2}\left( S_{+}^{13}S_{-}^{34}+h.c.\right) .
\label{H4eff}
\end{eqnarray}%
where%
\begin{eqnarray*}
\Phi (a^{\dagger }a,S^{jj}) &=&g_{1}\varepsilon _{1}\left( S^{22}\left(
S^{11}+1\right) +a^{\dagger }a\left( S^{22}-S^{11}\right) \right)  \\
&&+g_{2}\varepsilon _{2}\left( S^{33}\left( S^{11}+1\right) +a^{\dagger
}a\left( S^{33}-S^{11}\right) \right)  \\
&&+g_{3}\varepsilon _{3}\left( S^{44}\left( S^{22}+1\right) +a^{\dagger
}a\left( S^{44}-S^{22}\right) \right)  \\
&&+g_{4}\varepsilon _{4}\left( S^{44}\left( S^{33}+1\right) +a^{\dagger
}a\left( S^{44}-S^{33}\right) \right)
\end{eqnarray*}%
is the dynamic Stark shift \cite{stark}.

Let us classify the effective resonances present in the above (first
order)Hamiltonian:

\begin{enumerate}
\item \textit{Multiphoton resonances}:\textbf{\ }The only two resonances of
this type are one-photon and two-photon resonances. There are four
one-photon transitions corresponding to explicit resonances, which do not
appear in the effective Hamiltonian (\ref{H4eff}). The term $\sim
a^{2}S_{+}^{14}$ describes two-photon transitions, with corresponding
resonance condition $E_{4}-E_{1}\approx 2\omega $.

\item \textit{Virtual photon resonances:} the last four terms in (\ref{H4eff}%
), $\sim S_{+}^{24}S_{-}^{12}$, $S_{+}^{24}S_{-}^{13}$, $%
S_{+}^{12}S_{-}^{34} $, $S_{+}^{13}S_{-}^{34}$, with the resonance
conditions $E_{4}-E_{2}\approx E_{2}-E_{1}$, $E_{4}-E_{2}\approx E_{3}-E_{1}$%
, $E_{2}-E_{1}\approx E_{4}-E_{3},$ and $E_{3}-E_{1}\approx E_{4}-E_{3}$
respectively.

\item \textit{Photon assisted resonances: }The only resonance of this type
is of the form\textbf{\ }$\sim f\left( S^{44},S^{11},a^{\dagger }a\right)
S_{+}^{23},$ is produced in the first order of the perturbation theory
generating transitions between the middle atomic levels, and the
corresponding resonance condition is $E_{3}\approx E_{2}.$
\end{enumerate}

\section{Dynamic Resonances}

In this Section we study quantum systems corresponding to classically
non-integrable systems, mainly focusing on the models describing interaction
of atoms with quantized and classical fields without applying the Rotation
Wave Approximation \cite{sinrwa}, \cite{pla} (some different systems not
preserving the number of excitations are discussed in \cite{bombeo}).

It is convenient to start with a general analysis of such systems. The main
idea consists in removing the counter-rotating terms from the multi-channel
Hamiltonian (\ref{H1}) using a method of adiabatic elimination. In the limit
of weak interaction, $g_{jk}\ll \omega _{n},\Omega _{m}$ one can apply, for
instance, the Lie-transformation method outlined in the Appendix. Although,
such analysis can be performed for a general system, we will focus on the
simplest case of a one channel Hamiltonian and show, that even in this case
the absence of the integral of motion corresponding to the total number of
excitations leads to the appearance of a series of (dynamic) resonances,
which can be classified according the interaction type.

Consider a single-channel Hamiltonian (\ref{H1})%
\begin{equation}
H=\omega X_{0}+\Omega Y_{0}+g(X_{+}Y_{-}+X_{-}Y_{+})+g\left(
X_{+}Y_{+}+X_{-}Y_{-}\right) ,  \label{H2}
\end{equation}%
where $X_{0}$ ($Y_{0}$) is the free Hamiltonian of the subsystem $X$ ($Y$)
and $X_{+}$ ($Y_{+}$), $X_{-}$ ($Y_{-}$) are the up and down operators (see
Appendix), respectively, which describe transitions between the energy
levels in the subsystem $X$ ($Y$), they hold the $su_{d}(2)$ \cite{Karassiov}
commutation relations (\ref{escal}). From now on, we do not impose any
commutation relation between the transition operators, which are generally
some functions of the diagonal operators and integrals of motion $\left[
N_{1},X_{0,\pm }\right] =\left[ N_{2},Y_{0,\pm }\right] =0$,

\begin{eqnarray*}
\left[ X_{+},X_{-}\right] &=&P_{1}\left( X_{0}\right) =\nabla _{X_{0}}\phi
_{1}\left( X_{0},N_{1}\right) , \\
\left[ Y_{+},Y_{-}\right] &=&P_{2}\left( Y_{0}\right) =\nabla _{Y_{0}}\phi
_{2}\left( Y_{0},N_{2}\right) .
\end{eqnarray*}%
where $\nabla _{X_{0}}\phi _{1}\left( X_{0},N_{1}\right) $, $\nabla
_{Y_{0}}\phi _{2}\left( Y_{0},N_{2}\right) $ are given by the structural
functions (appendix (\ref{estruc})), and, in general, (we are going to omit
the dependence of the integral of motion)
\[
\nabla _{z}\phi \left( z\right) =\phi \left( z\right) -\phi \left(
z+1\right) .
\]

The consequences of the existence of the \textit{counterrotating} term $\sim
XY$ in the Hamiltonian (\ref{H2}) are: \ a) the dimension of the whole
representation space is the product of the corresponding dimensions of the
subsystems: $\dim (X\otimes Y)=\dim X\cdot \dim Y$, b) there are some
additional resonances apart from $\omega =\Omega $, that in the case of a
single-channel Hamiltonian (\ref{H3}) do not exist, c) The type of
resonances depends on the structure of algebras describing the $X$ and $Y$ \
systems.

The counterrotating\textit{\ }term rapidly oscillates (with frequency $%
\omega +\Omega $), and thus, can be eliminated by applying the Lie-type
transformation%
\begin{equation}
U=\exp \left[ \varepsilon \left( X_{+}Y_{+}-X_{-}Y_{-}\right) \right] ,\quad
\varepsilon =g/\left( \omega +\Omega \right) \ll 1,  \label{U}
\end{equation}%
to the Hamiltonian (\ref{H2}) and using the standard perturbative expansion%
\textbf{, }see Appendix . From now on we suppose that $\omega \geq \Omega .$

The elimination of the above term leads to the appearance of new elements in
the transformed Hamiltonian. All these new terms can be divided into three
groups: the first group contains the \textit{non-resonant terms,} of the
form $X_{+}^{n}Y_{+}^{m}+h.c.$, that can always be eliminated under the
condition $\omega ,\Omega \gg g$ by applying some suitable transformations;
the second group consists of \textit{resonant terms,} that cannot be removed
if certain relations between $\omega $ and $\Omega $ hold, since the
interaction becomes resonant ( the transformation which eliminates a given
term from the Hamiltonian becomes singular). This group contains terms of
the form $X_{+}^{n}Y_{-}^{m}+h.c.$, which describe transitions between
energy levels of the whole system. The third group includes the diagonal
terms (functions only of $X_{0},Y_{0}$), that can never be removed. Our
strategy consists of keeping in the Hamiltonian only diagonal terms and
resonant terms, and, we conserve only the leading order coefficients in
these terms.

All the counterrotating terms, like $\sim $ $X_{+}^{n}Y_{+}^{m},$ can be
eliminated by applying the Lie-type transformation
\begin{equation}
U=\exp \left[ \pi (\varepsilon _{1}..\varepsilon _{k})\eta \left(
X_{+}^{n}Y_{+}^{m}-X_{-}^{n}Y_{-}^{m}\right) \right] ,  \label{UmainT}
\end{equation}%
where $\eta \approx g/\left( n\omega +m\Omega \right) \ll 1$, and $\pi
(\varepsilon _{1}..\varepsilon _{k})$ is the product of powers of some small
parameters $\varepsilon _{j}\ll 1$, $j=1,....,k$. In fact $\pi (\varepsilon
_{1}..\varepsilon _{k})$ is proportional to the coefficient of the term $%
\sim X_{+}^{n}Y_{+}^{m}+X_{-}^{n}Y_{-}^{m}$. Applying a sequence of
appropriate transformations (\ref{UmainT}) we obtain the effective
Hamiltonian \cite{sinrwa}
\[
H_{eff}\approx \omega X_{0}+\Omega Y_{0}+g\varepsilon \Phi \left(
X_{0},Y_{0},\varepsilon \right) +
\]

\begin{equation}
+g\sum_{k=1}^{\infty }\sum_{l=0}^{\infty }\frac{(-\delta
)^{l+k-1}\varepsilon ^{l+2(k-1)}}{(k-1)!\left( l+k-1\right) !}\left[
X_{+}^{k}Y_{-}^{2l+k}\theta _{kl}(X_{0},Y_{0},\varepsilon )+h.c.\right] ,
\label{Heff}
\end{equation}%
where $\delta =g/2\Omega \ll 1.$

The term $\Phi \left( X_{0},Y_{0},\varepsilon \right) $ represents the
dynamic Stark shift (or Bloch-Siegert shift) \cite{bloch} and can be
expanded on powers of the small parameter $\varepsilon $ as follows
\begin{equation}
\Phi \left( X_{0},Y_{0},\varepsilon \right) =\phi _{1}\left( X_{0}\right)
\phi _{2}(Y_{0})-\phi _{1}\left( X_{0}+1\right) \phi
_{2}(Y_{0}+1)+O(\varepsilon ^{2}).  \label{phi}
\end{equation}%
The terms of the form$\ X_{+}^{k}Y_{-}^{2l+k}+h.c.$ describe all the
admissible resonant interactions. In particular the term with $k=1$, $l=0$
represents the principal resonance. The coupling constants $\theta
_{kl}(X_{0},Y_{0},\varepsilon )$ depend on the algebraic structure (\ref%
{estruc}) of the operators describing both subsystems (more precisely, on
the degree of the structural polynomials $\phi _{1}\left( X_{0},N_{1}\right)
$ and $\phi _{2}\left( Y_{0},N_{2}\right) $), and in particular, they can
become zero for some $k$, $l$ \cite{sinrwa}. The coefficient in the
principal term has the form
\[
\theta _{10}(X_{0},Y_{0})=1,
\]%
and

\begin{equation}
\theta _{kl}(X_{0},Y_{0})=\sum_{j=0}^{l+[k/2]}C_{l+[k/2]}^{j}\left(
P_{1}\left( X_{0}+k\right) \right) ^{l+[k/2]-j}\left( -P_{1}\left(
X_{0}\right) \right) ^{j}R_{k}(X_{0},Y_{0}-2j),  \label{theta}
\end{equation}%
\begin{eqnarray*}
R_{k}(X_{0},Y_{0}) &=&\sum_{j=0}^{[\frac{k-1}{2}]}C_{[\frac{k-1}{2}%
]}^{j}\left( P_{1}\left( X_{0}+k\right) \right) ^{[\frac{k-1}{2}]-j}\left(
-P_{1}\left( X_{0}\right) \right) ^{j}\nabla _{Y_{0}}^{k-1}\phi _{2}\left(
Y_{0}-2j\right)  \\
&&\times \prod\limits_{i=1-2j}^{k-2(j+1)}\phi _{2}\left( Y_{0}+i\right) ,
\end{eqnarray*}%
for $l\geq 1$, where $C_{k}^{j}$ are the binomial coefficients and $\nabla
_{z}^{0}f(z)=1$. Note that the product in the last equation is equal to
unity if the upper limit is less than the lower one.

In contrast to the case of kinematic resonances discussed in the previous
Section, the number of possible resonant interactions appearing in the
Hamiltonian (\ref{Heff}) is infinite. These interactions (\textit{dynamic
resonances}) can be classified as follows: a) Principal (explicit)
resonance, $\omega =\Omega ,$ corresponding to the interaction term that is
explicitly present in the original Hamiltonian $\sim X_{+}Y_{-}$; b)
Higher-order resonances: $\left( 2l+k\right) \Omega =k\omega $ where $l,k$ $%
\in N\quad $($l\geq 1,k\geq 1$), corresponding to the effective interactions
$\sim X_{+}^{k}Y_{-}^{2l+k}$, which can be divided into: b.1) odd
resonances: when $k=1,l=1,2,3,...$, that is $\omega =\left( 2l+1\right)
\Omega $, b.2) even resonances: $k=2,l=1,3,5...$, with the resonance
condition $\omega =2m\Omega ,\quad m=1,2,...$, b.3) fractional resonances: $%
\omega =\left( 2l/k+1\right) \Omega ,$ where $k$ and $l$ are coprime numbers.

It is worth noting that in the vicinity of each resonance $\left(
2l+k\right) \Omega =k\omega $, only the interaction term $\sim
X_{+}^{k}Y_{-}^{2l+k}$ survives. This means that if some resonance condition
is held, the system is in an approximate invariant subspace and there exists
an approximate integral of motion $N=\left( 2l+k\right) X_{0}+kY_{0}$.

This simple example shows that even in one-channel systems an infinite
number of resonant interactions may arise, if the total number of
excitations is not preserved. The appearance of such \textit{dynamic}
resonances in chaotic-like systems is expected from the point of view of
classical dynamic systems \cite{caos}. Nevertheless, the quantum nature of
interacting subsystems imposes certain restrictions on the possibility of
surviving of the dynamic resonances. We will discuss such restrictions for
two examples of the interaction of an atomic system with quantum and
classical fields.

\subsection{Atom-quantized field interaction}

Let us consider a collection of $A$ identical two-level atoms interacting
with a single mode of a quantized field (the Dicke model) without RWA. The
Hamiltonian that describes this system is%
\begin{equation}
H=\Omega a^{\dagger }a+\omega S_{z}+g(aS_{+}+a^{\dagger }S_{-})+g(a^{\dagger
}S_{+}+aS_{-}),  \label{Hdin1}
\end{equation}%
and the condition $\omega ,\Omega \gg g$ is held.

The following identifications
\[
X_{0}=S_{z},X_{\pm }=S_{\pm },\;Y_{0}=n,Y_{+}=a^{\dagger },Y_{-}=a,
\]%
so that $\phi _{1}(X_{0})=(1+A/2)A/2-X_{0}^{2}+X_{0},\phi _{2}(Y_{0})=Y_{0},$
lead to $\theta _{kl}\left( S_{z},n\right) =0,\,k\geq 3$ and the effective
Hamiltonian (\ref{Heff}) takes the form%
\begin{eqnarray}
H_{eff} &\approx &\omega S_{z}+\Omega a^{\dagger }a+g\varepsilon \left(
S_{z}^{2}+\left( 2a^{\dagger }a+1\right) S_{z}-(1+A/2)A/2\right)
\label{HDeff} \\
&&+g\sum_{l=0}^{\infty }\frac{\left( -2\delta \varepsilon \right) ^{l}}{l!}%
\left[ a^{2l+1}S_{+}+h.c.\right] -g\varepsilon \sum_{m=1}^{\infty }\frac{%
\left( 4\delta \varepsilon \right) ^{2m}}{(2m)!}\left[ a^{4m}S_{+}^{2}+h.c.%
\right] ,  \nonumber
\end{eqnarray}%
where $\varepsilon =g/\left( \Omega +\omega \right) ,\delta =g/2\Omega \ll
1. $

Note that the Hamiltonian (\ref{HDeff}) contains only the principal
(explicit) resonance ($l=0$), the even and the odd order interactions, but
no fractional resonances. This happens because the structural functions for
our subsystems are of a first degree polynomial of the photon number
operators $a^{\dagger }a$ and a second degree polynomial of the atomic
population operator $S_{z}$.

The integrals of motion for the series of odd and even (exact) resonances
are
\begin{eqnarray*}
N^{(kl)} &=&(2l+k)S_{z}+ka^{\dagger }a+2k\delta \left( S_{+}+S_{-}\right)
\left( a^{\dagger }+a\right) - \\
&&-\frac{2k^{2}\delta ^{2}}{l\left( l+k\right) }\left[ \left( 2l+k\right)
S_{z}\left( 2a^{\dagger }a+1\right) -kS_{z}^{2}\right] +O\left( \delta
{}^{3}\right) ,
\end{eqnarray*}%
where $l=2m+1$ in $N^{(2l)}$, $m=0,1,2,3,..$ and $k=1,2$ for even and odd
resonances correspondingly.

Let us recall that the higher resonances appear in the effective Hamiltonian
(\ref{HDeff}) only under the condition $\omega \geq \Omega $. In the
opposite case, $\omega \leq \Omega $, only the principal resonance survives
and in the approximation (\ref{Heff}) the whole effect of the
counter-rotating terms reduces to the dynamic Stark shift, which has the
same form as in the above case. This does not imply that there are no higher
resonances at all, but rather that they are essentially suppressed.

\subsection{Atom-classical field interaction}

We will proceed with the analysis of the interaction of atomic systems with
classical fields without RWA. A generic interaction Hamiltonian has the form%
\begin{equation}
H=\sum\limits_{jj}\omega _{j}X_{0}^{j}+\sum_{j}g_{j}\left(
X_{j}+X_{j}^{\dagger }\right) \cos (\Omega _{j}t+\vartheta _{j}),
\label{Hc1}
\end{equation}%
and admits some explicit resonances, $\omega _{j}=\Omega _{j}$. The whole
set of effective (implicit) interactions can be easily obtained in the same
way as it was outlined in 3.1. To be able to use the effective Hamiltonian
in the form (\ref{Heff}) we first rewrite the Hamiltonian (\ref{Hc}) in the
Floquet form by making use of the phase operators $\{E_{0j},E_{j}\}$, which
are generators of the Euclidean algebra:
\begin{equation}
\left[ E_{0j},E_{k}\right] =-\delta _{jk}E_{j},\quad \left[
E_{0j},E_{j}^{\dagger }\right] =\delta _{jk}E_{j}^{\dagger },\quad \left[
E_{j},E_{k}^{\dagger }\right] =0.  \label{Ej}
\end{equation}%
Each group of operators (labeled with the same index $j$) acts in a Hilbert
space spanned on the eigenstates of the (Hermitian) operator $E_{0j}:$%
\begin{equation}
E_{0j}|n\rangle _{j}=n|n\rangle _{j},\quad n=...-1,0,1,...,  \label{Ejes1}
\end{equation}%
so that in the basis (\ref{Ejes1}) the operators $E_{j},E_{j}^{\dagger }$
act as rising-lowering operators:
\[
E_{j}|n\rangle _{j}=|n-1\rangle _{j},\quad E_{j}^{\dagger }|n\rangle
_{j}=|n+1\rangle _{j}.
\]%
The phase states $|\vartheta \rangle _{j}$, the eigenstates of $E_{j},$ $%
E_{j}|\theta \rangle _{j}=e^{-i\vartheta _{j}}|\vartheta \rangle _{j}$, are
not normalized.

Now, let us consider the following time independent Hamiltonian
\begin{equation}
H^{op}=\sum_{j}\omega _{j}h_{j}+\sum_{jj}\Omega
_{j}E_{0j}+\sum_{j}g_{j}\left( X_{j}+X_{j}^{\dagger }\right)
(E_{j}+E_{j}^{\dagger }).  \label{HF}
\end{equation}%
It is easy to observe that the average value of the above Hamiltonian over
the phase states in the rotating frame,
\begin{eqnarray}
|\vartheta (t)\rangle  &=&\Pi _{j}e^{-i\Omega _{j}tE_{0j}}|\vartheta \rangle
_{j},  \label{Est} \\
|\vartheta \rangle _{j} &=&\lim_{N\rightarrow \infty }\frac{1}{\sqrt{2N+1}}%
\sum_{n=-\infty }^{\infty }e^{-i\vartheta _{j}n}|n\rangle _{j},  \nonumber
\end{eqnarray}%
where $|n\rangle _{j}$ are the basis states (\ref{Ejes1}), coincides with
the Hamiltonian (\ref{Hc1}) due to
\[
\langle \vartheta (t)|E_{0j}|\vartheta (t)\rangle =0,\quad \langle \vartheta
(t)|E_{j}|\vartheta (t)\rangle =e^{-i(\Omega _{j}t+\vartheta _{j})}.
\]%
The Hamiltonian (\ref{HF}) is a Floquet form of the initial time dependent
Hamiltonian (\ref{Hc1}) (from now on we put all the phases $\vartheta _{j}$
equal to zero). In the case of weak driven fields, $g\ll \omega ,\Omega $,
the Hamiltonian (\ref{HF}) can be represented in a form of expansion over
the principal resonances which can be observed in this system according to (%
\ref{Heff}).

In the single channel case, the Dicke model in the classical field, we have
the classical problem, first solved by Shirley \cite{shirley} (for a single
two level atom). Let us consider a collection of $A$ two level atoms in a
linearly polarized EM field. The corresponding Hamiltonian has the form
\begin{equation}
H=\omega S_{z}+g\left( S_{-}+S_{+}\right) \cos \Omega t,  \label{HDicke}
\end{equation}%
where $\Omega $ is the classical field frequency.

The Floquet form of (\ref{HDicke}) is
\begin{equation}
H^{op}=\omega S_{z}+\Omega E_{0}+g\left( S_{-}+S_{+}\right) (E+E^{\dagger }),
\label{Hop}
\end{equation}%
The following identifications
\[
X_{0}=S_{z},X_{\pm }=S_{\pm },\;Y_{0}=E_{0},Y_{+}=E^{\dagger },Y_{-}=E,
\]%
so that $\phi _{1}(X_{0})=(1+A/2)A/2-X_{0}^{2}+X_{0},\phi
_{2}(Y_{0})=E^{\dagger }E=EE^{\dagger }=I,$ lead to $\theta _{kl}\left(
S_{z},n\right) =0,\,k\geq 2$ which immediately leads to the effective
Hamiltonian (\ref{Heff})
\begin{equation}
H_{eff}^{op}=\omega S_{z}+\Omega E_{0}+2g\varepsilon E^{\dagger
}ES_{z}+g\sum\limits_{l=0}^{\infty }\frac{\left( -2\delta \varepsilon
\right) ^{l}}{l!}\left( E^{2l+1}S_{+}+h.c.\right) .  \label{HClass}
\end{equation}%
Averaging the above Hamiltonian over the phase states we obtain an effective
time-dependent Hamiltonian
\[
H_{eff}=\omega S_{z}+2g\varepsilon S_{z}+g\sum\limits_{l=0}^{\infty }\frac{%
\left( -2\delta \varepsilon \right) ^{l}}{l!}\left( e^{-i(2l+1)\Omega
t}S_{+}+h.c.\right) ,
\]%
which means that only odd resonances, $\Omega \approx \left( 2l+1\right)
\omega $, appear in this system. This result is different from the
atom-quantum field interaction (and surprisingly can not be obtained from
the corresponding quantum Hamiltonian (\ref{HDeff}) by substituting the
field operators by $c-$numbers), because in this case the structural
function (\ref{estruc}) for the Euclidean algebra (\ref{Ej}) is a constant,
that is, a zero-th degree polynomial function.

\section{Conclusions}

Quantum systems possessing as the integral of motion the operator
corresponding to the total number of excitations (classically integrable
systems) admit a simple classification of possible resonant transitions,
which are separated into explicit (which appear in the original
Hamiltonians) and implicit (effective) resonances. The total number of such
resonances is always finite and depends on the number of atoms and of atomic
levels. By a simple algebraic manipulation of the free Hamiltonian and the
integrals of motion, one can obtain some specific conditions for the
frequencies of interacting subsystems, which define the implicit resonances.
Nevertheless, the form of the interaction terms associated with each
implicit resonance condition can be found only after obtaining the effective
Hamiltonian (in a perturbative way).

Quantum systems not preserving the total number of excitations (classically
non-integrable systems) do not admit a simple classification of admissible
resonant transitions. The explicit form of the effective Hamiltonian is
needed to specify the allowed resonances, and their number is always
infinite. Such non-linear (on the generators of some Lie algebra)
Hamiltonians can be represented as a series of operators describing all the
possible transitions, which might become resonant under specific relations
between frequencies of interacting subsystems. The structure of the
effective Hamiltonian essentially depends on the algebraic structure of
interacting subsystems (polynomials $P_{X}\left( X_{0}\right) ,P_{Y}\left(
Y_{0}\right) $). In particular, it is reflected in the types of \nolinebreak
resonances which are allowed for a given system.

In the vicinity of each resonant transition all of the other transitions can
be considered as a perturbation. Thus, in the course of evolution some
specific finite-dimensional subspaces in the Hilbert space of the whole
system are approximately preserved, and for each of these invariant
subspaces there is a corresponding integral of motion.

\begin{acknowledgement}
This work is partially supported by the Grant 45704 of ConsejoNacional de
Ciencia y Tecnologia (CONACyT).
\end{acknowledgement}

\section{Appendix}

The method of Lie-type transformation (small rotations) \cite{SR}, \cite{JMO}
provides a \textit{regular} procedure for obtaining approximate Hamiltonians
describing the effective dynamics of nonlinear quantum systems. The idea of
this method is based on the observation that several quantum optical
Hamiltonians can be written in terms of polynomially deformed algebras $%
sl_{pd}(2,R)$ \cite{Karassiov}, \cite{Higgs79}, \cite{Rocek91}, \cite%
{Bonatsos93}, \cite{Sklyanin82}, \cite{Karasiov94},
\begin{equation}
H_{\mathrm{int}}=\Delta \ X_{0}+g\left( X_{+}+X_{-}\right) ,  \label{HintT}
\end{equation}%
where the operators $X_{\pm }$ and $X_{0}$ are generators of the deformed
algebra and satisfy the following commutation relations (\ref{escal}), and
\begin{equation}
\lbrack X_{+},X_{-}]=P(X_{0}),  \label{XpmT}
\end{equation}%
where $P(X_{0})$ is a polynomial function of the diagonal operator $X_{0}$
with coefficients that may depend on some integrals of motion $N_{j}.$ If $%
P(X_{0})$ is a linear function of $X_{0}$, the usual $sl(2,R)$ or $su(2)$
algebras are restored. If for some physical reason (depending on the
particular model under consideration) $\eta =g/\Delta \ll 1$ is a small
parameter, the Hamiltonian (\ref{HintT}) is \textit{almost} diagonal in the
basis of the eigenstates of $X_{0}$ and can be approximately diagonalized by
applying in a perturbative manner the following unitary transformation (a
\textit{small nonlinear rotation})
\begin{equation}
U=\exp \left[ \eta \left( X_{+}-X_{-}\right) \right] .  \label{U1}
\end{equation}%
Applying the transformation (\ref{U1}) to the Hamiltonian (\ref{HintT})
according to the standard expansion
\begin{equation}
e^{A}Be^{-A}=\sum_{k=1}^{\infty }\frac{\eta ^{k}}{k!}\mathrm{ad}_{A}^{k}(B),
\label{exp}
\end{equation}%
where $\mathrm{ad}_{A}$ is the adjoint operator defined as $\mathrm{ad}%
_{A}(B)=[A,B]$, we obtain
\begin{eqnarray}
H_{\mathrm{eff}} &=&UH_{\mathrm{int}}U^{\dagger }=\Delta \
X_{0}+g\sum_{k=1}^{\infty }\eta ^{k}\frac{k}{(k+1)!}\mathrm{ad}%
_{T}^{k}(X_{+}+X_{-}),  \label{ad} \\
&&  \nonumber
\end{eqnarray}%
where $T=X_{+}-X_{-}$ and we have taken into account that, due to (\ref{XpmT}%
),
\[
\mathrm{ad}_{T}(X_{0})=[T,X_{0}]=-\left( X_{+}+X_{-}\right) .
\]%
The effective Hamiltonian acquires the following form
\[
H_{\mathrm{eff}}=\Delta \ X_{0}+g\eta \sum_{k=0}^{\infty }{}^{\prime }\eta
^{k}\left[ X_{+}^{k}f_{k}(X_{0},\eta )+h.c.\right] ,
\]%
where $f_{k}(X_{0},\eta )$ is a function of the diagonal operator $X_{0}$
and can be represented as a series on $\eta $:
\[
f_{k}(X_{0},\eta )=\frac{2\left( k+1\right) }{\left( k+2\right) !}\nabla
^{k+1}\phi (X_{0})+\mathcal{O}(\eta ),
\]%
and
\begin{equation}
\phi (X_{0})=X_{+}X_{-}  \label{estruc}
\end{equation}%
is a structural function, $\nabla \phi (X_{0})=\phi (X_{0})-\phi
(X_{0}+1)=P(X_{0})$; the prime ( ${}^{\prime }$) in the above sum means that
the term with $k=0$ is taken with the coefficient $1/2$.

By keeping terms up to order $\eta $ we get
\begin{equation}
H_{\mathrm{eff}}=\Delta \ X_{0}+\eta g\nabla \phi (X_{0}),  \label{H1eff}
\end{equation}
and in the first approximation the resulting effective Hamiltonian is
diagonal on the basis of eigenstates of $X_{0}$.

The higher-order contributions always have the form $X_{+}^{k}f_{k}(X_{0})+%
\mathrm{h.c.}+g(X_{0})$. This makes the procedure of removing the
off-diagonal terms somehow trivial at each step, in the sense that it is
always obvious which transformation should be applied. For example, to
eliminate the terms of the form
\[
\eta ^{l}[X_{+}^{k}f_{k}(X_{0})+f_{k}(X_{0})X_{-}^{k}],
\]%
it suffices to apply the transformation
\begin{equation}
\exp (\frac{\eta ^{l+1}}{k}T_{k}),  \label{Tp}
\end{equation}%
with $T_{k}=X_{+}^{k}f_{k}(X_{0})-f_{k}(X_{0})X_{-}^{k}$, since the first
commutator of $\Delta \ X_{0}$ with $T_{k}$ cancels the corresponding term
in the Hamiltonian.

\end{document}